\documentclass{elsart}
\usepackage{graphicx,amssymb}
\journal{Solid State Communications}
\begin{document}
\begin{frontmatter}

\title{The effect of Cu substitution on the A$_{1g}$ mode of La$_{0.7}$Sr$_{0.3}$MnO$_3$ manganites}

\author[valencia]{G. De Marzi\thanksref{now}},
\thanks[now]{Present address: NMRC, Lee Maltings - Prospect Row, Cork, Ireland}
\ead{gdemarzi@nmrc.ie}
\author[well]{H. J. Trodahl},
\author[paris]{J. Bok},
\author[valencia]{A. Cantarero}, and
\author[valencia]{F. Sapi\~na}

\address[valencia]{ Materials Science Institute, University
of Valencia, PO Box 22085, 46071 Valencia, Spain}

\address[well]{
School of Chemical and Physical Sciences, Victoria University of
Wellington, PO Box 600, Wellington, New Zealand}

\address[paris]{ESPCI, 10 rue Vauquelin, 75231 Paris CEDEX 05, France}

\begin{abstract}


We report on the first Raman data of Cu substituted
La$_{1-y}$Sr$_{y}$Mn$_{1-x}$Cu$_x$O$_3$ ($0\leq x \leq 0.10$ and
$0.17\leq y \leq 0.3$, accordingly in order to have the same
Mn$^{4+}$/[Mn$^{4+}$+Mn$^{3+}$] ratio), collected in the frequency
range 100-900 cm$^{-1}$ and at room temperature, with parallel
($e_i\parallel e_s$) and crossed ($e_i\perp e_s$) polarizations of
the incident ({\em $e_i$}) and scattered ({\em $e_s$}) light.
Spectra were fitted with a Drude-Lorentz model, and peaks at
190-220 and 430 cm$^{-1}$, together with two broad structures
centered at near 500 and 670 cm$^{-1}$, have been found. We also
have observed that the A$_{1g}$ mode is substantially shifted with
increasing Cu substitution. The A$_{1g}$ phonon shift is a linear
function of the tolerance factor $t$ and the rhombohedral angle
$\alpha_r$, thus following the structural changes of the MnO$_6$
octahedra in the system.

\end{abstract}

\begin{keyword}
Raman spectroscopy \sep manganites \sep phonons \PACS 78.30.Hv
\sep 75.47.Ex \sep 63.20.Dj \sep 78.20.-e
\end{keyword}
\end{frontmatter}

\section{Introduction}

The intriguing physical properties and the potential technological
applications of $R_{1-x}A_x$MnO$_3$ pseudocubic manganites (where
$R$ is a rare-earth metal: La, Pr, Nd, Dy; and $A$ is an alkaline
earth: Sr, Ca, Ba, Pb) have led to a resurgence of interest in
these and related materials. In particular, the phase diagram of
these compounds is complex, and many variables such as pressure,
applied magnetic field, doping concentration {\em x}, and $A$-site
average ionic radius can determine a wide range of ground states
in the system~\cite{Millis}. In many of these systems a
metal-insulator transition (MI) and colossal magnetoresistance
occur near the Curie temperature T$_c$, when the system undergoes
a paramagnetic (PM) to ferromagnetic (FM) transition. To explain
the PM to FM phase transition and the changes in transport
mechanisms, Zener proposed the so-called ``double exchange (DE)''
model~\cite{Zener}. In the framework of the DE model, two carriers
hop simultaneously, one from a Mn$^{3+}$ ion to an adjacent
O$^{2-}$ site and the other from the O$^{2-}$ to a neighboring
Mn$^{4+}$. However, numerous experimental results suggest that
other considerations influence the phase transitions, among them
the electron-phonon interaction arising from Jahn-Teller (JT)
distortions~\cite{Millis2}, the orbital degree of
freedom~\cite{Maezono}, the electron-electron correlations and the
coupling between spin and orbital structure~\cite{Ishihara}. The
JT lattice distortion in particular is thought to be larger in the
PM phase (above T$_c$), but to decrease~\cite{Millis2}
significantly after the transition to the FM phase. Thus, the
MnO$_6$ octahedra are highly distorted for T$>$T$_c$ but the
magnitude of this distortion decreases as T approaches T$_c$. The
idea is that a MI transition can occur more easily if the
octahedra are undistorted and the Mn--O--Mn angles tends to
180$^\circ$~\cite{Hwang,Garcia}; this effect can also be obtained
by changing the average dimension of the atom at the $A$ and/or
$B$ site. Studies of $R_{1-x}A_x$MnO$_3$ with $x$ close to 30\%
have shown the existence of a relationship between T$_c$ and the
distortion of the perovskite structure as measured by the
tolerance factor, {\em t}, or the mean size of cations at the A
site $\langle r_A\rangle$~\cite{Hwang,Maignan}. More precisely, it
has been shown that in perovskite series with a fixed $\langle
r_A\rangle$ value, the T$_c$ depends on the disorder at the $A$
site~\cite{Rodriguez,Fadli}; this disorder is quantified in terms
of the variance of the $A$ cationic radii distribution,
$\sigma^2(\langle r_A\rangle)$. There is substantially less
information in literature regarding the substitution in the B
cationic sublattice. Some attention has been paid to the
substitutions on the charge ordered ground state (i.e., 50\%
Mn$^{4+}$)~\cite{Maignan2}, but there are few results concerning
Mn-substitution having the optimum content of Mn$^{4+}$ (ca.
30\%). There is not yet enough data to correlate the dependence of
T$_c$ with any structural variable involving $B$-site
substitution~\cite{Gayathri}. The size of the substitutional
defects is a key factor influencing transition
temperatures~\cite{Fadli}; the structural disorder produces a
strong local stress in MnO$_6$ octahedra (resulting in a
rotation), modifying the Mn--O--Mn angles and thus changing
lattice and electronic properties. Thus both vibrational and
electronic properties of mixed valence manganites with defect in
the $B$ cation sublattice are influenced by structural disorder
introduced by local internal pressure associated to the
substitutional defects. In this paper we present a Raman
scattering investigation on the manganese perovskites
La$_{1-y}$Sr$_y$Mn$_{1-x}$Cu$_x$O$_3$ as a function of $B$-site Cu
doping. The copper-free end member of the series,
La$_{0.7}$Sr$_{0.3}$MnO$_3$ (LSMO), has been extensively studied
with a great variety of techniques; for spectroscopic spectra, see
for example optical conductivity
measurements~\cite{Okimoto,Quijada,Takenaka}, and Raman
scattering~\cite{Gupta,Podobedov,Amelitchev,Granado}.
Nevertheless, reflectivity, Raman and/or ellipsometric studies
have not been yet performed on
La$_{1-y}$Sr$_y$Mn$_{1-x}$Cu$_x$O$_3$. In this paper we report a
Raman study of a series containing a fixed ratio
Mn$^{4+}$/(Mn$^{4+}$+Mn$^{3+}$) = 0.3.

\section{Experiment}

Substitution in specific sites in the lattice affects the
vibrational modes through changes to the mass, the bonding
strength and the bonding configuration, so that Raman spectroscopy
and infrared (reflectance and ellipsometry techniques) are  ideal
tools for studying the role of $B$-site substitution in
manganites. Moreover, Raman spectroscopy is a useful probe for
those materials in which lattice, as well as electron dynamics,
strongly interacts by means of electron-phonon interaction. With
this in mind we have measured Raman spectra of polycrystalline
La$_{1-y}$Sr$_y$Mn$_{1-x}$Cu$_x$O$_3$ samples, searching for
effects due to the changes in both electron-phonon coupling and
free carrier screening in the metallic phase. As an ease to
interpretation the values of $y$ are selected to keep constant the
ratio Mn$^{4+}$/(Mn$^{4+}$+Mn$^{3+}$) = 0.3 through the series.
Details on the preparation of the studied samples, together with
x-ray diffraction and magnetization measurements are described
elsewhere~\cite{Fadli}, while values of $y$ and Cu content, T$_c$
temperature, tolerance factor, and rhombohedral angle are given in
Table I. Raman spectra were collected at room temperature with
back-scattering geometry of the incident laser light from surfaces
polished down to 1 $\mu$m with diamond paste. The samples were
excited with the 514.5 nm line of a Ar ion laser and the scattered
light analyzed using a triple monochromator system (DILOR XY800)
with a LN$_2$ cooled Charge-Coupled device (CCD) detector. The
power was kept below 15 mW focused in a spot diameter of 0.1
mm$^2$, in order to avoid heating the samples. Spectra were
collected with parallel ($e_i\parallel e_s$) and crossed
($e_i\perp e_s$) polarizations of the incident ({\em $e_i$}) and
scattered ({\em $e_s$}) light. In Figs. 1 and 2, respectively, the
Raman spectra collected in the parallel-polarized and
cross-polarized configurations are shown.

\section{Results and discussion}

As in the case of previously published spectra for
La$_{0.7}$Sr$_{0.3}$MnO$_3$~\cite{Amelitchev,Granado}, and many
related materials~\cite{Pantoja}, the most striking features are
two anomalously broad lines centered near 500 and 670 cm$^{-1}$.
In addition to these very broad structures, there are narrower
lines at 190-220 and 430 cm$^{-1}$. It should be noticed that the
shoulder at around 600 cm$^{-1}$ disappears for $x = 0$. In the
cross polarized configuration, the first peak at about 200
cm$^{-1}$ completely disappears. The Raman spectra showed in Fig.
1 were fitted in the 130-900 cm$^{-1}$ region frequency with the
following expression for the Raman intensity~\cite{Yoon}:
\begin{equation}
S(\omega)=\frac{A(\Gamma/2)^2}{\omega^2+(\Gamma/2)^2}+
\sum_{i=1}^{5}
\frac{A_i\omega_i\omega\gamma_i^2}{(\omega^2-\omega_i^2)^2
+(\omega\gamma_i)^2}\quad,
\end{equation}
where the first term represents the electronic contribution to the
spectrum, consisting of ``collision-dominated'' low-frequency
response associated with diffusive hopping of the carriers. The
second term represents the ``odd lorentzian'' shape of a phonon
mode ($\omega_i$, $\gamma_i$, and $A_i$ are the peak frequency,
linewidth, and amplitude, respectively). The error on the
frequency resulting from the fit analysis is always smaller than
the resolution of the spectrometer (1.5 cm$^{-1}$ when the
aperture of the slit is 200 $\mu$m). Therefore, we assume that the
error is 1.5 cm$^{-1}$ for all the frequencies. We found four
peaks at 180-215, 430, 530, and 670 cm$^{-1}$. Note that a high
frequency oscillator (near 1100 cm$^{-1}$) is necessary to achieve
a good fit. It should be pointed out that this High Frequency
Contribution (HFC) is also found in Raman experiments on
La$_{0.75}$Ca$_{0.25}$MnO$_3$~\cite{Congeduti}, and it has been
suggested that the HFC should be associated with the
photoionization of small polarons. But we argue that this is a
very clear evidence that the feature up there is second order
phonon scattering.  In fact, the HFC structure is also present in
Sr doped manganites, and all other pseudocubic and double layer
manganites~\cite{Congeduti,Pantoja2} and we believe that such
features are weak second-order Raman scattering effects (density
of state), mirroring the bands seen in the region 400-800
cm$^{-1}$. The frequencies of the experimental peaks are plotted
in Fig. 3, wherein it is evident that the narrow lower peak shows
a substantial shift as function of Cu concentration. Our samples
exhibit rhombohedral crystal structure~\cite{Fadli}, space group
$D^6_{3d}(R_{\overline{3}c})$. According to site group analysis,
30 vibrational modes are predicted:
\begin{equation}
\Gamma(D^6_{3d})=2A_{1u}+3A_{2g}+A_{1g}+4A_{2u}+4E_g+6E_u\quad,
\end{equation}
where the $A_{1g}+4E_g$ modes are Raman active, the $3A_{2u}+5E_u$
are infrared active, and the $A_{1u}+3A_{2g}$  are silent modes.
Following the procedure described in Ref. \cite{Porto}, the La (or
Sr) ions participate in four phonon modes
($A_{2g}+A_{2u}+E_g+E_u$), the Mn (or Cu) ions take part in four
phonon modes ($A_{1u}+A_{2u}+2E_u$), while the oxygen ions
contribute in twelve modes
($A_{1g}+A_{1u}+2A_{2g}+2A_{2u}+3E_g+3E_u)$. As a consequence, it
is found that the $A_{1g}$ modes only involve the motion of the
oxygen ions in the $C_2$ sites, while the $E_g$ arise both from
oxygen site and La (Sr) site vibrations. Following Granado {\em et
al.}~\cite{Granado}, we assign the lowest frequency peak at
190-220 cm$^{-1}$ (M1) to an $A_{1g}$ mode, and the peak at 430
cm$^{-1}$ (M2) to a $E_g$ symmetry. The assignment of the M1 peak
to an $A_{1g}$ mode is confirmed by the fact that this mode is
absent in cross polarized configuration, as it is expected under
these conditions (we remind the selection rules for this mode:
$A_{1g} \to \alpha_{xx}+\alpha_{yy}, \alpha_{zz}$,  see Ref.
\cite{Porto}). Concerning the peaks at 530 cm$^{-1}$ (M3) and 670
cm$^{-1}$ (M4), Granado {\em et al.} suggest that the M3 and M4
peaks should be due to scattering induced by orthorhombic
distortions which may be present on the surface. On the other hand
these modes are very broad in all manganites, which is an evidence
that they are full-band modes, rendered Raman active by disordered
Jahn-Teller distortions~\cite{Pantoja,Pantoja2}. The M1 mode
corresponds to a rotation of the octahedra around the hexagonal
{\em c-axis}~\cite{Granado}, which is closely related with the
$A_g$ mode in the orthorhombic ({\em Pnma})
structure~\cite{Amelitchev} (an {\em in-phase rotation} of the
MnO$_6$ octahedra around the {\em b-axis)}. Concerning the
frequency shift of the $A_{1g}$ mode, this does not simply involve
the motion of the La (Sr) ions: in fact, in such a case the
frequency should decrease as the heavier La ions are substituted
for Sr (see the values for {\em y} in Table I): actually, what is
seen is exactly the opposite behavior. It is worth noting that,
according with the site group analysis previously discussed, the
$A_{1g}$ mode involves only the motion of the oxygen ions in the
$C_2$ sites. Therefore it is reasonable to relate the changes on
the $A_{1g}$ mode frequency to the modifications of the oxygen
octahedra. Moreover, the introduction of substitutional defect in
the $B$-site (like Cu) has a strong effect in the structural
changes of the lattice. Since Cu substitution induces a strong
local stress, it can be expected that MnO$_6$ octahedra rotate,
and the Mn-O bond lengths decrease under this compression. Also
the Sr substitution may introduce structural disorder, but it
should be noticed that, since the Sr and La ions have similar
radii, the strength of the distortion induced by those ions is
less important. In conclusion, we believe that the observed
frequency shift with doping is determined by the distortions
introduced by the large substitutional Cu ions, involving the
motion of the oxygen cages and an effective rotation of the
octahedra. In order to be more quantitatively, let us discuss the
tolerance factor ({\em t}), which is defined as:
\begin{equation}
t=\frac{d_{A-O}}{\sqrt{2}d_{B-O}}=\frac{\langle r_A\rangle
+r_O}{\sqrt{2}(\langle r_B\rangle +r_O)}\quad ,
\end{equation}
where {\em d} denotes the interatomic distance simply calculated
 as the sum of ionic medium radii;  $\langle r_A\rangle$ is the average ionic
radius of $A$-site ions (La$^{3+}$ and Sr$^{3+}$) and $\langle
r_B\rangle$ is the average radius of the $B$-site ions (Mn and
Cu). The tolerance factors is $t=1$ for spherical ions packed in
the ideal perovskite structure (Mn--O--Mn angle equal to
180$^\circ$, and no-tilt of the octahedra) and becomes smaller as
$\langle r_B\rangle$ is increased by Cu substitution and the
lattice becomes more distorted; this means that the Mn--O--Mn
angle deviates from 180$^\circ$, which leads to a reduction of the
DE mechanism. The correlation between the frequency shift of the
A$_{1g}$ mode and the tolerance factor (calculated using the
Shannon radii \cite{Shannon}) is shown in Fig. 4. Within the
errors, it can be concluded that the scaling of the A$_{1g}$ mode
with the tolerance factor is linear. It is important to notice
that the changes in the tolerance factor have been induced by both
A- and B-site substitution; in literature, there is substantial
previous Raman data on the manganites as a function of A-site
substitution: an interesting work by Amelitchev {\em et al.}
\cite{Amelitchev} showed that the band shift of the A$_{1g}$ mode
is a linear function of the tolerance factor for a fixed value of
{\em x} (like 0.3 or 0.45). In their work, the tolerance factor is
changed by introducing different cations (La, Nd, Sr, Ca, Pr, Sm,
Eu, Gd) in the A sublattice of the perovskite structure. In order
to compare the effect of A-site and both A- and B-site
substitutions, we have added in Fig.4 their results for
R$_{0.7}$A$_{0.3}$MnO$_3$ (R = La$_{1-x}$Pr$_x$, A = Ca, Sr,
Sr$_{0.5}$Ca$_{0.5}$), R$_{0.67}$Sr$_{0.33}$MnO$_3$ (R =
Nd$_{0.5}$Sm$_{0.5}$, Sm, Eu, Gd), and
R$_{0.55}$Sr$_{0.45}$MnO$_3$ (R = Nd, Nd$_{0.5}$Sm$_{0.5}$, Sm,
Eu, Gd). The solid lines are a linear fitting of
R$_{0.7}$A$_{0.3}$MnO$_3$ ($\omega$ = 2975-2995$\cdot$t) and
R$_{0.55}$Sr$_{0.45}$MnO$_3$ ($\omega$ = 3088 - 3092$\cdot$t)
\cite{Amelitchev}. Our data are in good agreement with the
R$_{0.7}$A$_{0.3}$MnO$_3$ series; in our case, $\omega$ = 2853 -
2864$\cdot$t, and linear extrapolations to higher t suggest that
the mode should soften to zero at t = 0.996 (t = 0.993 and t =
1.001 for R$_{0.7}$A$_{0.3}$MnO$_3$ and
R$_{0.55}$A$_{0.45}$MnO$_3$). In facts, the linear scaling of the
A$_{1g}$ mode with tolerance factor is independent of the means by
which the tolerance factor is changed, as long as the
Mn$^{3+}$/Mn$^{4+}$ ratio is kept at a fixed value.
 Moreover, it is worth noting that
the frequency of the $A_{1g}$ mode can also be correlated with the
rhombohedral angle $\alpha_r$, whose deviation from 60$^\circ$
defines a measure of the rhombohedral distortion with respect to
the ideal cubic structure. The values for $\alpha_r$ have been
obtained from x-ray diffraction data, and are reported in Table I.
The linear relationship between the frequency shift of the
$A_{1g}$  mode and $\alpha_r$ is shown in the inset of Fig. 4. It
can be concluded that the band shift is a linear function of the
tolerance factor and the rhombohedral angle $\alpha_r$, that is,
the $A_{1g}$ mode follows the structural changes of the oxygen
cages.

\section{Conclusions}

Our experimental studies of the Raman spectra of
La$_{1-y}$Sr$_y$Mn$_{1-x}$Cu$_x$O$_3$ reveal that two strong peaks
appear at 190-220 and 430 cm$^{-1}$, together with two broad
structures centered at near 500 and 670 cm$^{-1}$. It is found
that the lowest peak (assigned as an $A_{1g}$ mode) shifts
linearly, with the increase of Cu concentration, toward higher
frequencies as {\em x} increase. Within an analysis in terms of
tolerance factor and of rhombohedral angle ($\alpha_r$) of the
$D^6_{3d}(R_{\overline{3}c})$ structure, it is proved that the
shift is determined by the structural disorder introduced by
substitutional Cu ions in the $B$-site, e.g., the shift follows
the changes of the MnO$_6$ octahedra in the system.

Finally, it should be pointed out that the structural distortions
can be influenced by the size of the B-site ion.  According to the
Shannon radii (r$_M$), the Cu ionic radius (r$_{Cu^{2+}}$ = 0.730)
is relatively large compared to the Mn radius (r$_{Mn^{3+}}$ =
0.645, r$_{Mn^{4+}}$ = 0.530); in contrast, ions such Cr$^{3+}$,
Co$^{3+}$, Ga$^{3+}$ (r$_{Cr^{3+}}$ = 0.615, $r_{Co^{3+}}$ =
0.610, $r_{Ga^{3+}}$ = 0.620) are similar to the Mn mean ionic
radius ($r_{<Mn>}$ = 0.611, calculated for a mean oxidation state
of 3). Therefore, Raman measurements on (La,Sr)MnO$_3$ systems
with smaller B-site substitutional defect (such as Cr$^{3+}$,
Co$^{3+}$, Ga$^{3+}$) are highly desirable: according to our
interpretation, in this case the phonon frequency shift should be
less pronounced.

\section{Acknowledgements}

The authors would like to thank Prof. N. Joshi, Dr. S. Lupi and
Dr. A. M. Power for useful discussion. This work was supported by
the UE through a RTN grant (HPRN-CT2000-00021).

\newpage

\newpage

\begin{table}[htbp]
\begin{center}
\begin{tabular}{c c c c c }
\hline
Cu content & {\em y} doping & T$_{c}$ & tolerance factor & $\alpha_r$\\
\hline \hline
0.00 & 0.300 & 372 & 0.929983 & 60.42202 \\
0.02 & 0.274 & 358 & 0.928021 & 60.45980 \\
0.04 & 0.248 & 331 & 0.926063 & 60.50238 \\
0.06 & 0.222 & 308 & 0.924109 & 60.55491 \\
0.08 & 0.196 & 274 & 0.922160 & 60.59863 \\
0.10 & 0.170 & 236 & 0.920216 & 60.64173 \\
\hline
\end{tabular}
\end{center}
\caption{ Chemical analysis, observed T$_{c}$(K), tolerance
factor, and rhombohedral angles $\alpha_r$(degrees), for samples
of nominal composition La$_{1-y}$Sr$_{y}$Mn$_{1-x}$Cu$_x$O$_{3}$.
}
\end{table}

\newpage

    \begin{figure}
    \begin{center}
    \includegraphics{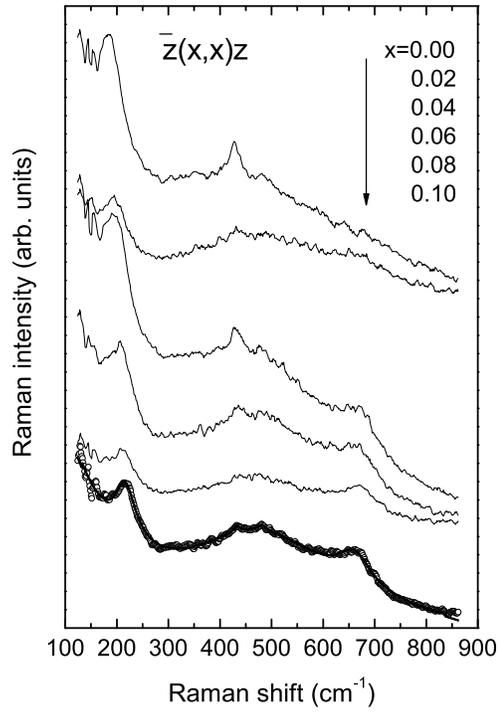}
    \end{center}
    \caption{Room temperature Raman spectra in parallel polarized geometry for
     polycrystalline La$_{1-y}$Sr$_{y}$Mn$_{1-x}$Cu$_{x}$O$_{3}$ samples. The solid line represents the fitting to Eq.1.}
    \end{figure}

\newpage
    \begin{figure}
    \begin{center}
    \includegraphics{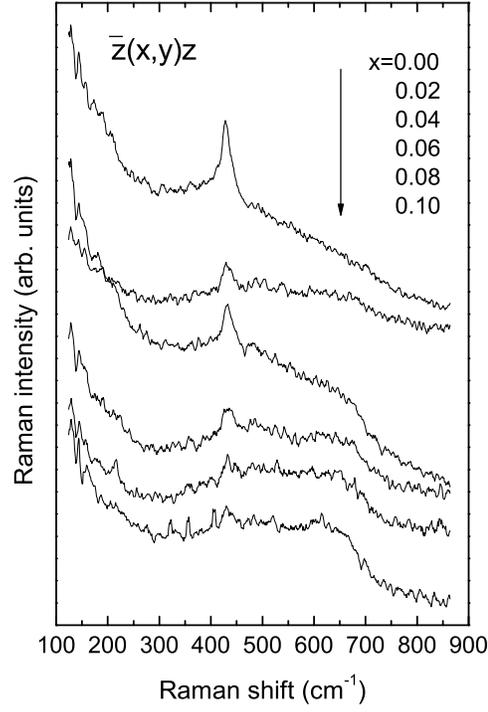}
    \end{center}
    \caption{Room temperature Raman spectra in cross polarized geometry for
     polycrystalline La$_{1-y}$Sr$_{y}$Mn$_{1-x}$Cu$_{x}$O$_{3}$ samples. }
    \end{figure}

\newpage
     \begin{figure}
     \begin{center}
     \includegraphics{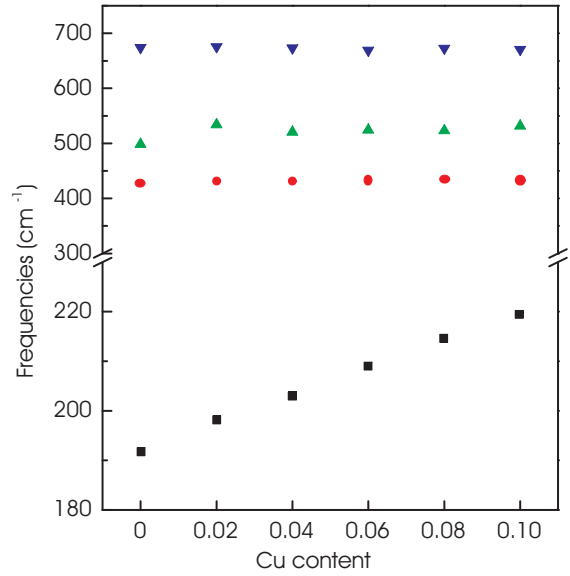}
     \end{center}
     \caption{$B$-site doping dependence of the four Raman shift peaks, as obtained by Eq. (1).}
     \end{figure}

\newpage
    \begin{figure}
    \begin{center}
    \includegraphics{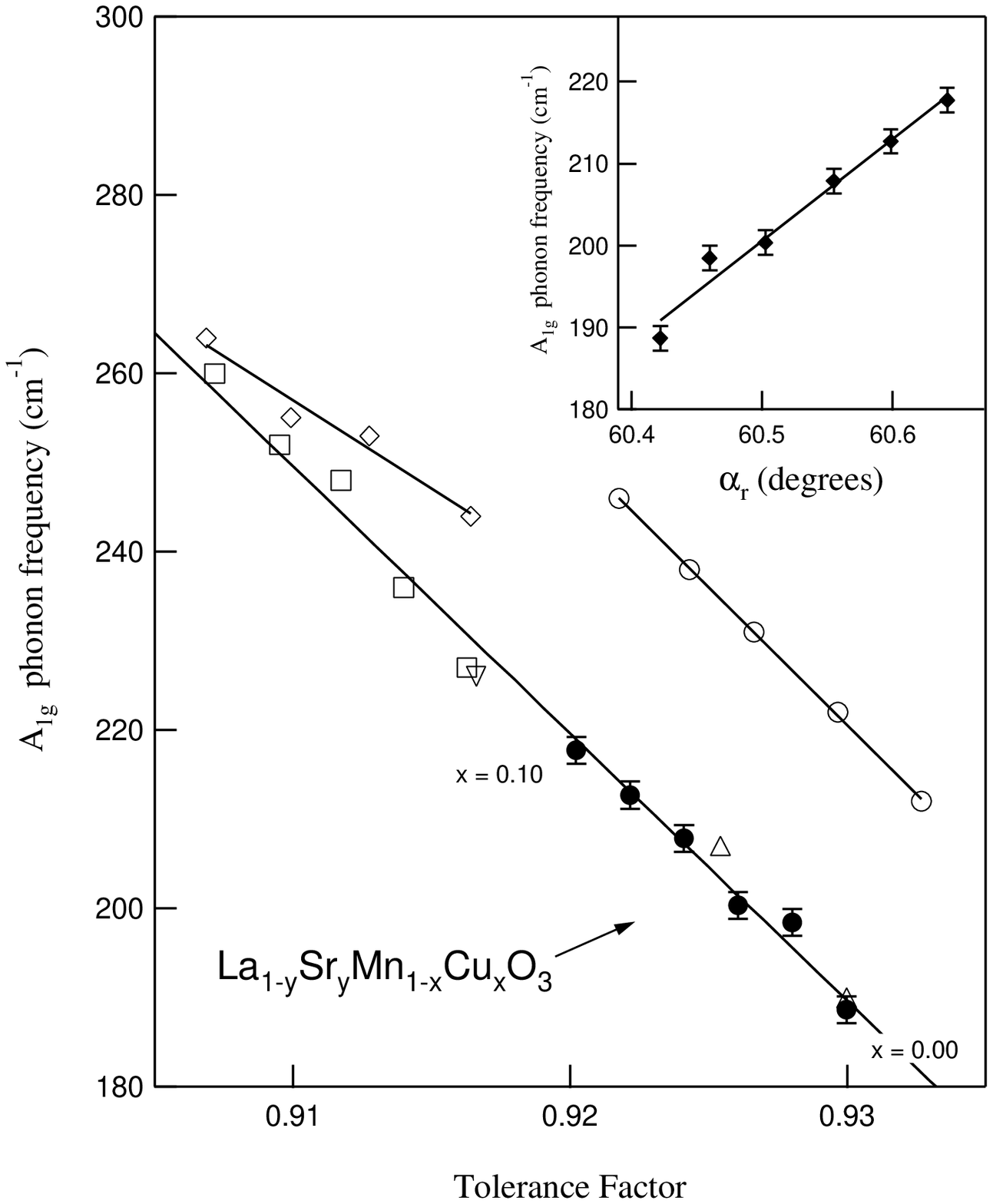}
    \end{center}
    \caption{Plot of the A$_{1g}$ frequency shift as a function of the tolerance
factor {\em t} for La$_{1-y}$Sr$_y$Cu$_x$Mn$_{1-x}$O$_3$(filled
circles). The following data from ref.18 are also shown in the
figure: La$_{1-x}$Pr$_x$Ca$_{0.3}$MnO$_3$ ( $\Box$ ; x = 0, 0.25,
0.5, 0.75, 1), La$_{1-x}$Pr$_x$Sr$_{0.3}$MnO$_3$ ( $\triangle$ ; x
= 0, 0.5),
(La$_{0.5}$Nd$_{0.5}$)$_{0.7}$(Sr$_{0.5}$Ca$_{0.5}$)$_{0.3}$MnO$_3$
( $\bigtriangledown$ ), R$_{0.67}$Sr$_{0.33}$MnO$_3$ ( $\Diamond$
; R = Nd$_{0.5}$Sm$_{0.5}$, Sm, Eu, Gd), and
R$_{0.55}$Sr$_{0.45}$MnO$_3$ ( $\circ$ ; R = Nd,
Nd$_{0.5}$Sm$_{0.5}$, Sm, Eu, Gd). The solid lines are a linear
fitting of R$_{0.7}$A$_{0.3}$MnO$_3$ ($\omega$ =
2975-2995$\cdot$t) and R$_{0.55}$Sr$_{0.45}$MnO$_3$ ($\omega$ =
3088-3092$\cdot$t)\cite{Amelitchev}. A linear fit to our data (not
shown in the figure) gives $\omega$ = 2853-2864$\cdot$t. In the
inset, the A$_{1g}$ frequency shift is plotted as function of the
rhombohedral angle $\alpha_r$.}
\end{figure}

\end{document}